# Sub-Doppler optical resolution by confining a vapour in a nanostructure

Philippe BALLIN, Elias MOUFAREJ, Isabelle MAURIN,
Athanasios LALIOTIS, Daniel BLOCH*

*Laboratoire de Physique des Lasers, Université Paris 13, Sorbonne Paris-Cité
CNRS, UMR 7538, 99 Avenue J-.B. Clément, F-93430 Villetaneuse, France*

## ABSTRACT

We show that a thermal vapor confined in a nanostructure is of spectroscopic interest. We perform reflection spectroscopy on a Cs vapour cell whose window is covered with a thin opal film (typically, 10 or 20 layers of ~ 1µm diameter spheres). Sub-Doppler structures appear in the optical spectrum in a purely linear regime of optical excitation and the signal is shown to originate from the interstitial regions of the opal. These narrow spectral structures, observable for a large range of oblique incidence angles (~ 30-50°), are an original feature associated to the 3-D vapor confinement. It remembers a Dicke narrowing, *i.e.* a Doppler broadening suppression when the atomic motion is sub-wavelength confined. This narrowing, commonly observed in the r.f. domain when a buffer gas ensures a collision confinement effect, had remained elusive in the optical frequency. Also, we describe preliminary experiments performed in a pump-probe technique, intended to elucidate the spatial origin of the narrow contribution. We finally discuss how our results allow envisioning micron-size references for optical frequency clocks, and high resolution spectroscopy of weak and hard-to-saturate molecular lines.

## 1. INTRODUCTION

An atomic or molecular vapour is the simplest material providing a universal frequency reference, in the radio and optical domain as well. The corresponding frequency reference systems (clocks) are generally bulky, because the atoms must be kept far away from the surface, and a low atomic density has to be chosen to minimize atom collisions, implying big volumes. With the blossom of communication technologies, there is an urgent need to make frequency reference devices as compact as possible[1]. Recent advances demonstrate that highly confined vapors still allow performing atomic or molecular physics, notably with sealed, meter-long, photonic fibers[2]. The interstitial regions of a porous media have also been used for spectroscopy[3]. However, the Doppler broadening, which causes a major broadening, is not eliminated with these techniques, except when applying to the holey fiber cell the saturated absorption technique[4], a technique restricted to rather strong transitions and sensitive to transit time broadening.

With the reduction in size, various extra benefits can be expected. In particular, confinement is susceptible to reduce the Doppler broadening, through the Dicke narrowing[5], *i.e.* when the atomic free path is typically shorter than the wavelength. This narrowing, well-known in the *r.f.* domain, when buffer gas collisions restrict the atomic free motion, remains elusive in the optical domain, as a considerable buffer gas density would be required at optical wavelengths.

---

* e-mail address : *daniel.bloch@univ-paris13.fr*

## 2. EXPERIMENTS WITH GAS CONFINED IN AN OPAL

Here, we report on experiments performed with a dilute alkali-metal (Cs) atomic vapor, confined in the interstitial regions of an opal made of glass nanospheres (diameter d ~ 1μm, occasionally d ~ 400 nm). We use glass cells, sealed with Cs vapour inside. A thin opal has been deposited on one of the window of the cell by a layer-by-layer Langmuir-Blodgett technique[6]. To avoid the formation of Cs clusters, an adequate thermal control of the system has to be implemented. Optically, the thin opals are highly scattering media, and the transmission is weak (for 10 layers of d = 1 μm spheres, it is below 1% at the wavelengths $\lambda_{D1}$ = 894nm, $\lambda_{D2}$ = 852 nm of Cs resonance). In spite of the corrugated nature of the interface, one observes specular reflection from the opal-covered window (fig.1), so that one can monitor the reflection coefficient when tuning the laser frequency across the atomic resonance (reflection spectroscopy). In addition, the sensitivity of the reflection spectrum to the searched narrow contribution is enhanced by a frequency derivation technique (FM applied to the laser, and detection of the modulated part of the signal).

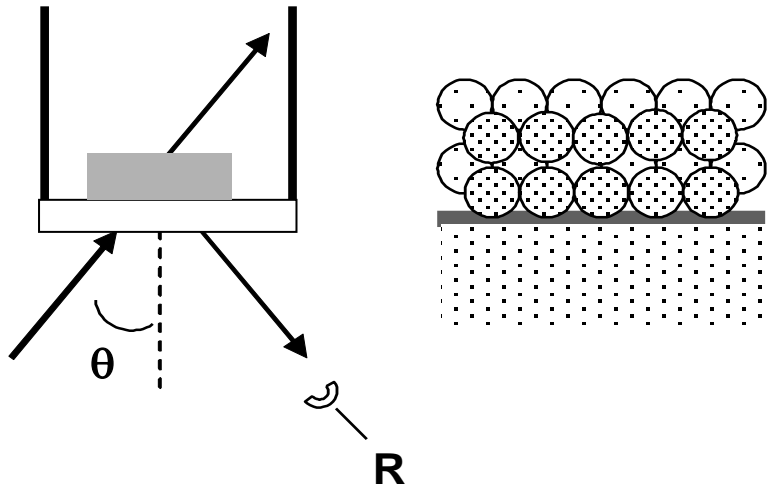

Fig.1. *Scheme of the experimental set-up (left): opal in grey; and zoom of an opal structure deposited on a flat window (right)*

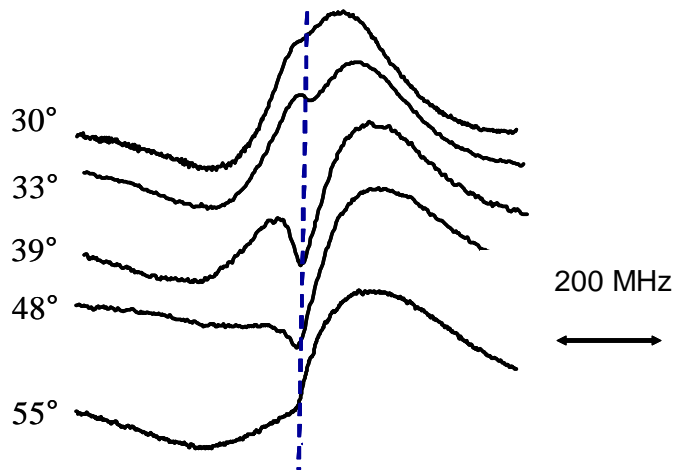

Fig.2: *FM reflection spectrum as a function of the external incidence angle at the interface ($D_1$ Cs line, 10 layers of 1 μm opal), in TM polarization. The dashed line indicates the center of an auxiliary saturated absorption spectrum.*

Our major observations are the following:
i) Under nearly normal incidence, the opal reflection lineshape exhibits a sub-Doppler resonance 30 MHz width, *vs.* an expected Doppler width exceeding 200 MHz
(ii) The lineshape broadens up when increasing the incidence angle and the sub-Doppler structure becomes less apparent
(iii) For large incidence angles (external angle ~ 30-60°) a narrow sub-Doppler structure ($\leq$ 30 MHz) appears (see fig.2), superimposed to the broad Doppler-broadened structure, whose shape evolves with the incidence angle. The narrow structure is precisely located at the Doppler-free resonance (as shown by comparison with a frequency marker). This appearance of a narrow contribution is better seen for the TM polarization. It is also possible to nearly isolate it (*i.e.* broad background suppressed) through a second harmonic FM detection.

Another remarkable point, important for applications, is that this sub-Doppler structure, not resembling any previously reported optical signal, appears in the linear regime, *i.e.* the sub-Doppler features do not require a minimal level of irradiating intensity to be observed. A saturation regime can be reached for the opal-confined vapor, but for an intensity much larger (~20 times) than in the ordinary vapor (a situation where optical pumping to an hyperfine ground state level can take place). This higher onset for this saturated regime is related to the shorter interaction time for confined atoms.

### 3. DISCUSSING THE ORIGIN OF THE NARROW STRUCTURE

Until now, only very few techniques are able to generate sub-Doppler and *linear* spectroscopic signals with a vapour: only selective reflection spectroscopy at a flat window/vapour interface[7], which is a technique probing the medium on a typical depth $\lambda/2\pi$, and thin cell spectroscopy[8], produce such signals under the condition of an irradiation under normal incidence. In both techniques, different transient regimes are experienced for atoms arriving to, or departing from, the surface (*i.e.* behavior dependent upon the normal velocity). This normal axis is the same one on which the Doppler effect is counted when the irradiation is under normal incidence, allowing a response selecting "slow" atoms. Here, the sub-Doppler signal found at normal incidence may indicate a similar effect, but the sub-Doppler contribution found at large incidence angles is a feature with no equivalent.

This is why we discuss below the spatial origin of the signal:

- First, the observed results are similar for the 10 and 20 layers opal, in spite of the stronger scattering of the 20 layers opal, that attenuates the transmission through the opal. This establishes that the contribution of the opal/vapor interface, analogous to selective reflection at a corrugated interface, can be neglected.

- Second, one needs to note that in an opal, one has to distinguish the first half-layer, which is a region of a lower density, from the periodical *compact* arrangement following the first equatorial plane of the spheres (see fig.1). Usually, only this periodical arrangement is considered when an opal is described as a photonic crystal. Moreover, calculations show that the nearly empty first (half-)layer seems to be the main responsible for the (non resonant) specular reflection observed with the cell. This nearly empty first (half-)layer could be responsible for a behavior resembling the one of a vapor nanocell (*i.e.* 1-D confinement)[8], justifying a sub-Doppler behavior under normal incidence and the broadening with the incidence angle. Conversely, it cannot explain the narrow structure observed for large incidence angles. In this last case, a more in-depth propagation of the field (to be described through near-field optics, as d and $\lambda$ are comparable) can justify that the narrow resonant signal is associated to a probing of atoms confined inside the small interstitial regions of the opal (of a typical size d/2). The sensitivity of the lineshape to polarization (fig.3) confirms that different atomic responses are explored when changing the polarization behavior, showing that propagation *inside* the opal is an important feature. This is confirmed when comparing $D_1$ and $D_2$ lines by the observation of differences in the mixture between broad and sub-Doppler responses that can be traced back to differing propagations. Indeed, even a small change of the $\lambda$/d ratio modifies notably the propagation features (*e.g.* transmission), notably at large angles. Again, this difference indicates that the propagation of light inside the opal largely differs for the two wavelengths.

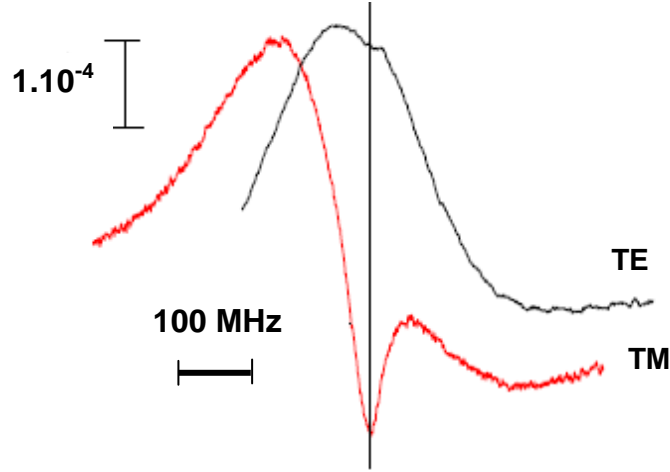

Fig.3: *Comparison between the refection spectrum ($D_1$ line) at θ = 40 ° for a TE and TM polarized beam. The vertical scale corresponds to the relative intensity change, the vertical mark is a (saturated absorption) frequency reference.*

The narrow contribution at large incidence angles, attributed to the interstitial regions of the opal, is reminiscent of the Dicke narrowing, To interpret this observation of a Dicke narrowing in the optical domain, we underline major differences between buffer gas collisions, and collisions with the hard glass spheres of the opal. Collision of the atom with a glass sphere leads to a genuine energy quenching, so that the atom rebuilds its interaction with light from the ground state. Conversely, for an optically excited atom, collision with a buffer gas essentially brings an out-phasing of the optical dipole, rather than an effective energy de-excitation, that could be only carried through a transfer to the (relatively weak) kinetic energy.

For a deeper interpretation, an exact theory describing reflection lineshapes at an opal interface would be needed. It is however a heavy work, as it requires to calculate the transient resonant behavior of a distribution of atoms located in the interstitial regions. Even if this interaction can probably be modelled with a mean-field theory, a description of the field structure inside the opal interstices is clearly needed. Such a task (that we are presently pursuing) has not been addressed in the current literature, and remains a complex problem in nanooptics.

## 4. COMPLEMENTARY PUMP-PROBE EXPERIMENTS and LIAD

Alternately, one may expect to gain information on the spatial origin of the narrow Dicke-type contribution in a pump-probe experiment, in which a spatially selective (pump-induced) optical pumping would enable to probe specific layers in the opal. The information to be gained may remain limited, because in its principle, such an experiment should still require that the propagation for the pump beam is well-known. However, a strong pump irradiation is susceptible to wash out all of the pump-illuminated atoms, so that playing with the pump beam incidence could offer enough versatility to gain sensitive information on the origin of the observed narrow peaks on the probe beam. In addition, the pump beam effect should be more related with its transmitted intensity, rather than with its more complex reflective behavior.

We have implemented such experiments with the pump and probe beams delivered by two different lasers. Usually, we scan the probe beam frequency across one of the hyperfine manifold of the Cs $D_2$ line (852 nm), while the pump beam is tuned at a fixed frequency, in the vicinity of the $D_1$ line (894 nm). An AM is applied to the pump beam (chopper), and only the modulated part of the probe beam is detected. Such a dual-beam set-up allows a large variety of experiments, in particular with respect to the incidence angle of each beam, or to its polarization. The results that we have obtained are nearly insensitive to the pump polarization; this is in agreement with the idea that the pump efficiency is connected to transmission of the pump beam, which is much less sensitive to polarization than reflection. Also, in all cases, one observes a relative enhancement of the narrow contribution appearing on the probe beam spectrum, as long as this narrow structure can be observed (weak incidence angles, or a range ~30-60 °). However, the incidence angle range for

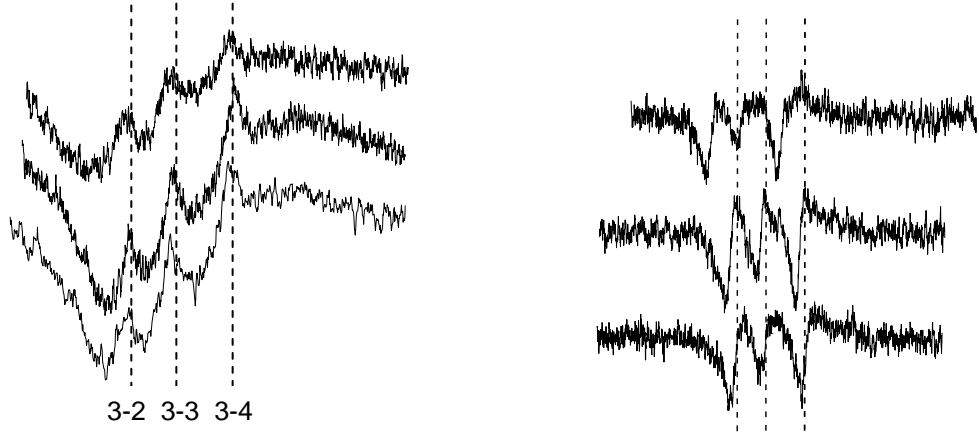

         3-2  3-3  3-4

Fig.4 *Comparison between an AM pump-probe spectrum (scan of the probe frequency) with a probe incidence +58 °, and a pump incidence of 2° (left), or -58° ( right). The vertical lines correspond to the positions of the Doppler-free transitions, as marked by an auxiliary saturated absorption. The bottom spectra are for a pump exactly resonant with the $D_1$ line, the middle ones for a ~ - 30 MHz shift, the upper ones for a ~ - 200 MHz shift -comparable to the Doppler width- .*

which a narrow structure can be observed, is not sensitively extended by using this pump-beam geometry. Moreover, we find that for the situation of the largest interest, *i.e.* of a probe beam sent under an oblique incidence -*e.g.* 40 °, as in fig. 3- , the narrow peaks remain centered onto the Doppler-free transition when the pump beam is sent under a normal incidence, whatever is the pump frequency. Surprisingly, if the pump beam is sent an incidence comparable to the one of the probe beam (we have only worked in a coplanar geometry, mostly with nearly opposite incidence angles), the position of the narrow peaks on the probe spectrum depends on the pump beam frequency (see fig.4). It happens in a way rather comparable to a velocity selective pump-probe experiment, well-known when an axial velocity selection is operated, as obtained with pump and probe beams freely (counter- or co-) propagating along a same axis, or even with evanescent waves; however, a major difference is that here, no cross-over resonances appear.

For these pump-probe experiments, it has always been needed to tune the pump frequency on a transition leaving the same ground state hyperfine level than the one on which absorption is probed. This confirms that, inside the opal there is not enough time for a ground state population transfer through an optical pumping process. To check this with more details, we have tried to increase our discrimination of pumped atoms by a temporal analysis, varying the frequency of the amplitude modulation applied to the pump beam. We have found that even for extremely slow modulation frequencies (the set-up did not allow us to go below 10 Hz, still the behavior at 10 Hz was very different than at 100 Hz), a phase shift still occurs between the applied modulation, and the one detected on the probe. The corresponding time constant is extremely long compared to any atomic trajectories within the free regions of the opal, and is long even compared to the expected values of Cs atom sticking time on a glass surface. We suspect that a Light-Induced Atomic Desorption (LIAD) process[9] takes place in the glass nanospheres, whose intimate structure[10] could be porous.

## 5. CONCLUSION

The major result is the observation of a sub-Doppler signal in linear spectroscopy for a large range of incidence angles. A sensitive issue for applications, requiring further experimental studies, is the control of the shape of the narrow structure, which apparently mixes-up symmetric and dispersion-like features, depending upon the incidence angle.

Presently, our work allows envisioning very compact sub-Doppler references, based upon weak (and hence hardly saturable) molecular lines, as a benefit of the intrinsic linearity of the method combined with the Dicke narrowing. The probed volume yielding a sub-Doppler response can be conceptually as small as several spheres (*i.e.* several $\lambda^3$), and the irradiation extremely focused. Alternate to the reflection measurements reported here, a detection of scattering may also be considered, allowing complex or irregular frontiers of the confining nanostructured medium. One may hence expect, with just a slightly larger volume, that an efficient multiple scattering will increase the effective optical path inside the resonant vapor, as observed recently[3]. Another extension of our experiments would be to look for an equivalent sub-Doppler contribution in a porous media, when the average pore size is $\leq \lambda$. Finally, let us note that as long as the confinement length remains above ~100 nm, the atom-surface interaction [7,8] perturbs only marginally the atomic resonances, as shown by our present experiments. In the same manner, the atomic motion should tend to wash out the photonic effects associated to a bandgap structure of the periodically organized confining medium.

## ACKNOWLEDGEMENTS

Work performed in the frame of the ANR project Mesoscopic gas 08-BLAN-0031, and bilateral programmes ECOS-Sud U08E01 (with Uruguay) and CNRS PICS #5813 (with Russia). The opal was prepared at CRPP-Bordeaux by S. Ravaine team.